\documentclass[aps,pra,twocolumn,groupedaddress,showpacs]{revtex4-1}
\usepackage{graphicx}

\newcommand{\beq}{\begin{equation}}
\newcommand{\eeq}{\end{equation}}
\newcommand{\bea}{\begin{eqnarray}}
\newcommand{\eea}{\end{eqnarray}}
\newcommand{\bra}[1]{\langle #1|}
\newcommand{\ket}[1]{|#1\rangle}
\newcommand{\braket}[2]{\langle #1|#2\rangle}

\begin{document}


\title{Quantum-enhanced Phase Estimation with an Amplified Bell State}


\author{Jaspreet Sahota}
\email[]{jsahota@physics.utoronto.ca}

\author{Daniel F. V. James}

\affiliation{CQIQC and IOS, Department of Physics, University of Toronto, 60 Saint George Street, Toronto, Ontario M5S 1A7, Canada}


\date{\today}

\begin{abstract}
We propose a phase estimation protocol for optical interferometry that employs a probe state (containing on average $\bar{n}$ photons) obtained by squeezing each mode, separately, of a single photon path entangled Bell state. This scheme involves a Mach-Zehnder type interferometer for which each mode is squeezed after the first beam splitter. Information about the differential phase is extracted using a parity detection and the resulting measurement signal is super-resolving and supersensitive, with a minimum phase uncertainty $\Delta \varphi = 2/(\bar{n}+1)$. This probe state can be generated with current technologies where $\bar{n}$ is in the order of many thousands of photons. 
\end{abstract}



\maketitle



\section{Introduction}

\begin{figure}[b]
\centering
\includegraphics[width=6.6 cm]{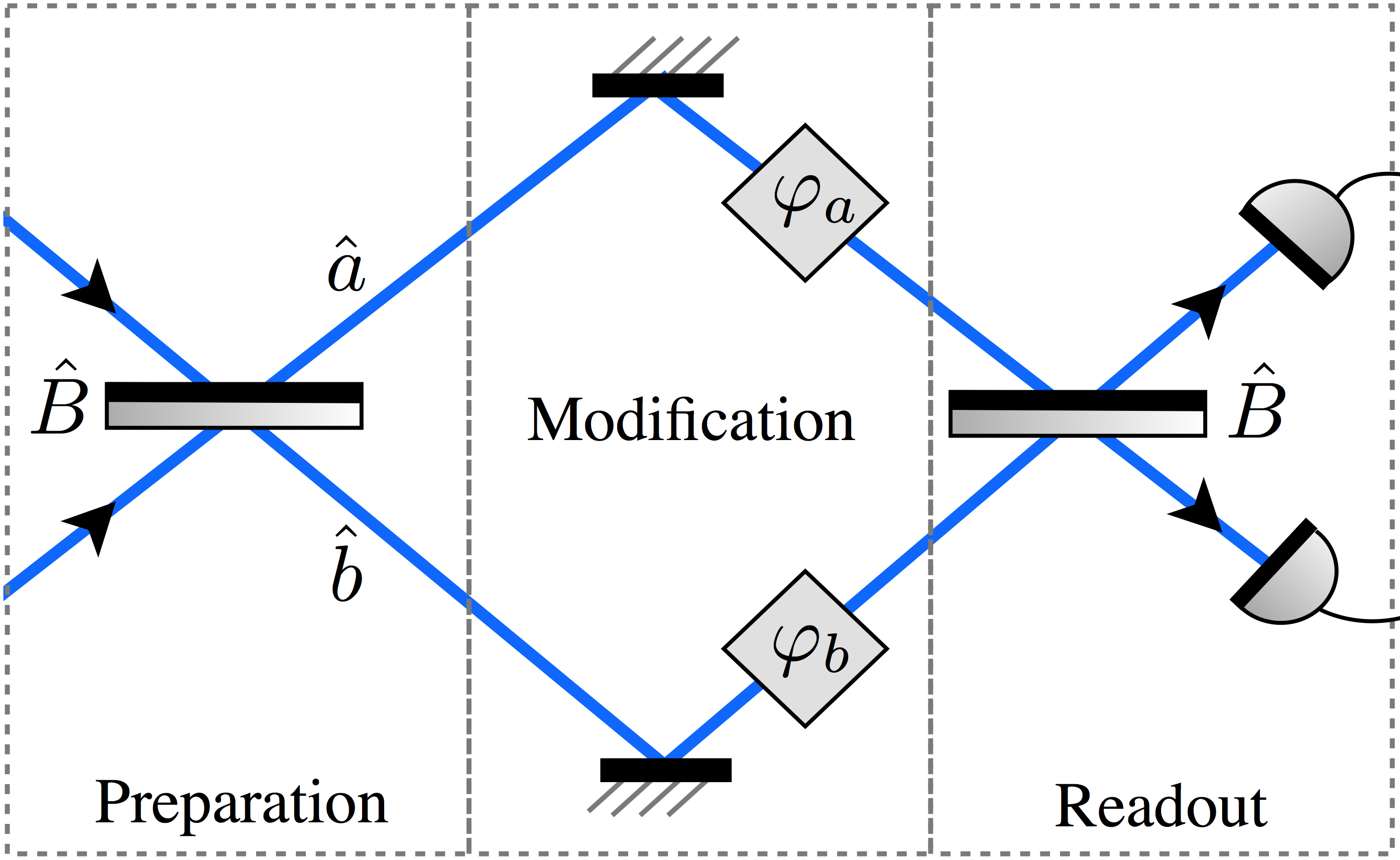}
\caption{(Color online)  The three steps of parameter estimation in Mach-Zehnder interferometry. Probe preparation: An initial field is injected into a BS, resulting in output modes $\hat{a}$ and $\hat{b}$. Probe modification: These optical modes acquire a phase difference $\varphi=\varphi_a -\varphi_b$, resulting from their corresponding path difference. Probe readout: We recombine these modes with a second BS and perform a photon counting measurement on the output.}
\label{a}
\end{figure}

The most sensitive measurements in the fundamental and applied sciences are typically made with optical and atomic interferometers. This includes optical interferometry for detecting gravitational waves \cite{LIGO11}, Ramsey interferometry for measuring properties of atoms and molecules \cite{RevModPhys.81.1051}, and optical lithography for nano-device fabrication \cite{Pavel2013259}. The Mach-Zehnder interferometer (MZI) is a paradigm for these technologies (Fig. \ref{a}). Thus, improvements in phase resolution and sensitivity of MZIs have profound implications on the scientific disciplines of imaging, sensing, and information processing. 

The two optical paths inside the MZI acquire a differential phase $\varphi={\varphi_a -\varphi_b}$ via a linear interaction, which can be estimated with an error $\Delta\varphi \ge 1/\sqrt{\bar{n}}$ (the lower bound is referred to as shot-noise limit or the standard quantum limit) when a laser beam, containing on average $\bar{n}$ photons, is injected into the primary input port of the initial beam splitter (BS). The use of non-classical light to enhance optical interferometry performance has a long history, starting in 1981 when Caves discovered that injecting a squeezed vacuum into the secondary input port of the initial BS leads to sub-shot-noise error \cite{Caves81}. 

A quantum advantage can be obtained in metrology by using certain quantum states of light (probe states) that contain non-classical correlations \cite{Giovannetti:2011ys}. For example, the maximally path entangled noon state, $\ket{noon}=(\ket{n,0}+\ket{0,n})/\sqrt{2}$, is a $n$-photon probe state that attains the so-called Heisenberg limit $\Delta\varphi = 1/n$ under ideal conditions \cite{PhysRevA.61.043811, PhysRevLett.85.2733, PhysRevA.66.013804, doi:10.1080/0950034021000011536}. This is the ultimate lower limit on $\Delta\varphi$ when an $n$-photon probe state is employed and the phase $\varphi$ is acquired via a linear interaction \cite{PhysRevLett.75.2944,PhysRevLett.77.2352, PhysRevLett.96.010401, PhysRevLett.105.180402, PhysRevA.85.041802, PhysRevLett.108.210404}.  Much work has focused on developing quantum schemes that can realize this limit \cite{PhysRevLett.56.1515, PhysRevA.33.4033,PhysRevLett.71.1355,Jacobson:1995fk,PhysRevA.56.R1083, PhysRevLett.100.073601, PhysRevLett.110.163604,PhysRevA.61.043811, PhysRevA.66.013804, doi:10.1080/0950034021000011536}.

The practical implementation of noon states into phase estimation protocols is in part hindered by their extreme sensitivity to photon loss \cite{Gilbert:08, PhysRevA.80.063803}. This has initiated efforts to find $n$-photon probe states that are optimal in the presence of photon loss \cite{PhysRevLett.102.040403, PhysRevA.83.021804, Escher:2011zr, escher2011quantum, PhysRevA.78.063828}. In addition, it has been an extremely daunting task to generate noon states in the laboratory for large $n$ \cite{PhysRevLett.87.013602, Steinberg04, Kim:09, Afek14052010, PhysRevLett.107.163602}, with the current record being $n=5$ \cite{Afek14052010}. As a result, classical (or laser) interferometry, which operates above the shot-noise limit in practice, still provides significantly superior phase resolution than noon state metrology of current standards.   

In another prevailing approach to quantum-enhanced interferometry, the requirement of fixed total photon number is relaxed and the MZI is injected with the brightest available non-classical states of light (large $\bar{n}$), that have in general a fluctuating photon number, in an attempt to attain resolution competitive with classical interferometry \cite{Caves81, PhysRevLett.110.163604, PhysRevLett.104.103602}.

In this paper we propose an alternative approach  to quantum-enhanced phase estimation based on elementary operations from quantum optics: The passive optical elements that make up a standard MZI are used and active optical elements (spontaneous parametric down-converters) will be required to prepare the probe state. We start by injecting a MZI with a single photon state to produce a path entangled Bell state. This state is then amplified to a macroscopically entangled probe state inside the interferometer. The resulting probe state, like the noon state, exists in a superposition of two macroscopically distinct parts. That is, one part of the superposition represents the majority of the photon being in one optical arm and the other part of the superposition represents the majority of the photons being in the other optical arm of the MZI. We show that our protocol has a minimum phase error of $\Delta\varphi_{\text{min}}= 2/\left(\bar{n}+1\right)$ when a parity measurement is utilized. The main advantage of this strategy is that current technologies allow for the production of such probe states with $\bar{n} \approx 10^4$ \cite{PhysRevLett.100.253601}. 

\section{Phase Estimation}

The problem of parameter estimation in quantum metrology is comprised of three steps. In the first step, we prepare a quantum state $\ket{\Psi}$ that will serve as the probe. Next, we let this probe state interact with a system of interest such that $\ket{\Psi}$ evolves to a state $\ket{\Psi_{\varphi}}$ which depends on the parameter $\varphi$ that we would like to estimate. In the final step, we measure an observable $\hat{O}$ of the modified probe $\ket{\Psi_{\varphi}}$ in order to gain information about the value of $\varphi$. The error in this measured quantity, given by the standard deviation $\Delta \hat{O} = \sqrt{\langle  \hat{O}^2 \rangle - \langle  \hat{O} \rangle^2 }$, propagates to the error in $\varphi$ as given by \cite{PhysRevA.33.4033}
\beq
\Delta^2 \varphi =  \frac{\Delta^2 \hat{O}}{|{\partial \langle  \hat{O} \rangle}/{\partial \varphi}|^2} \label{error},
\eeq
which is the ratio of the signal noise to how the signal changes with respect to $\varphi$. 

In a MZI experiment (Fig. \ref{a}), the probe is prepared by injecting an initial state $\ket{\Psi_0}$ into a 50:50 BS, which results in the output $\hat{B} \ket{\Psi_0} = \ket{\Psi}$, where the BS is described by the unitary operator $\hat{B}= \exp{\left[i\frac{\pi}{4}(\hat{a}^{\dag} \hat{b} + \hat{b}^{\dag} \hat{a})\right]}$. The output optical paths of this BS, which correspond to modes $\hat{a}$ and $\hat{b}$, acquire a differential phase $\varphi = \varphi_a - \varphi_b$. This modification of the probe state is described by the unitary $\hat{U}= \exp( i \varphi_a \hat{a}^{\dag} \hat{a} + i \varphi_b \hat{b}^{\dag} \hat{b})$. Finally, the optical paths of the modified probe $\ket{\Psi_{\varphi}} = \hat{U} \ket{\Psi}$  are recombined with a second 50:50 BS, resulting in a final state that is measured by performing a particular photon counting measurement depending on the nature of $\ket{\Psi}$.

In general, it can be very difficult to determine the observable $\hat{O}$, which minimizes (\ref{error}); however, we can use the quantum Fisher information (QFI) to determine the theoretical lower bound to this error. The QFI is an upper bound to the classical Fisher information optimized over all conceivable measurements that can be performed on $\ket{\Psi}$ in order to gain information about the value of $\varphi$. This quantity, defined as   
\beq
\mathcal{F}=4 \left( \braket{\Psi_\varphi^{\prime}}{\Psi_\varphi^{\prime}} - |\braket{\Psi_\varphi^{\prime}}{\Psi_\varphi}|^2 \right), \label{QFI}
\eeq   
where the primes denote derivatives with respect to $\varphi$, gives us the optimal error bound on $\Delta \varphi$:
\beq
\Delta^2 \varphi \ge \frac{1}{\mathcal{F}}, \label{CRB}
\eeq 
called the quantum Cram\'{e}r-Rao bound \cite{1976quantum, Caves94}.

\begin{figure}[b]
\centering
\includegraphics[width=6.3 cm]{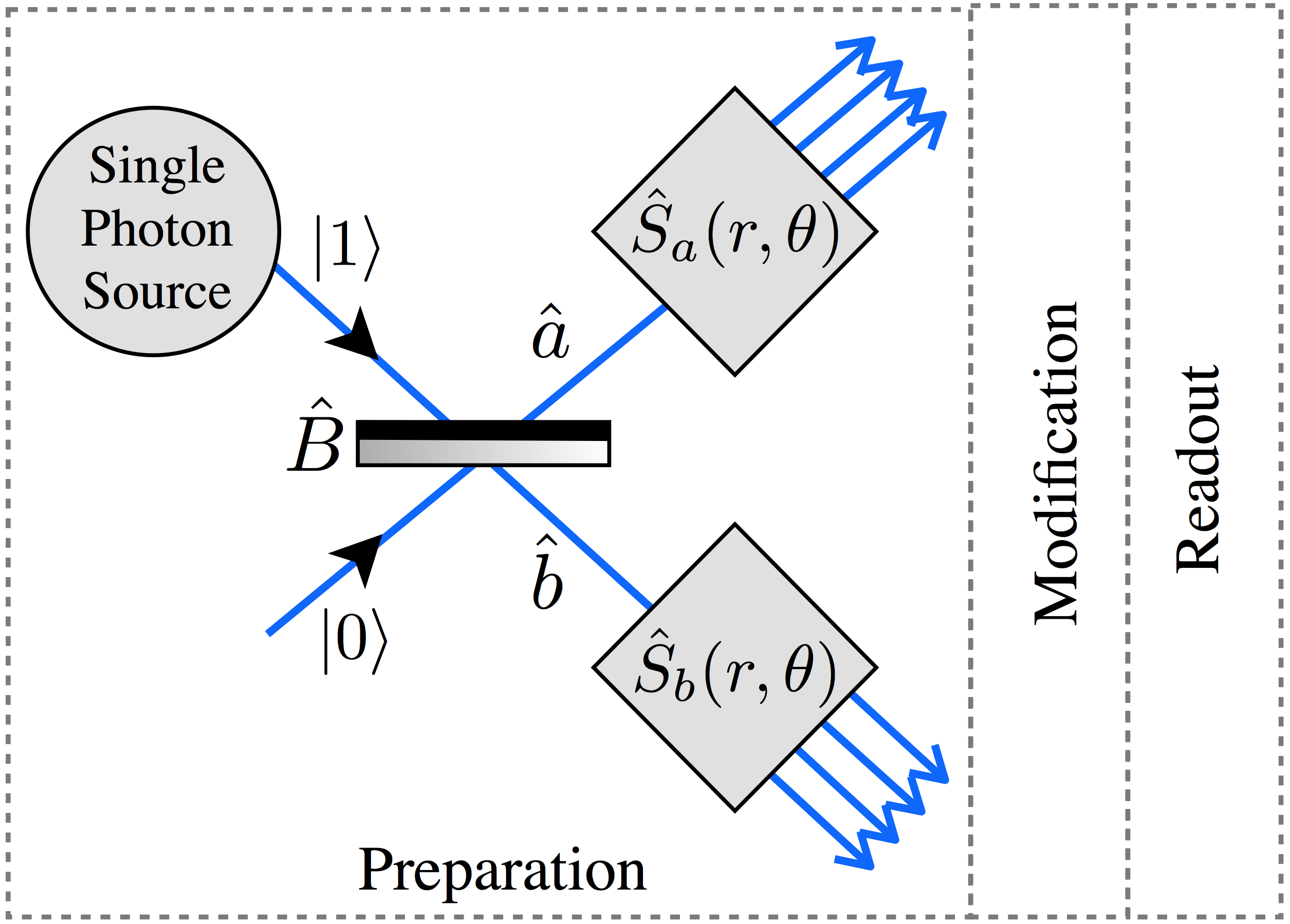}
\caption{(Color online) Schematic diagram showing the probe preparation step of our protocol. The modification and readout steps are as depicted in Fig.\ref{a}.}
\label{b}
\end{figure} 

It will be convenient for the following discussion to use the Schwinger representation \cite{PhysRevA.33.4033},
\bea
\hat{J}_1 = \frac{1}{2}\left(\hat{a}^{\dag} \hat{b} + \hat{b}^{\dag} \hat{a} \right), &\,&  \, \hat{J}_2 = - \frac{i}{2}\left( \hat{a}^{\dag} \hat{b} - \hat{b}^{\dag} \hat{a} \right) \nonumber \\
\hat{J}_3 = \frac{1}{2} &\,& \left( \hat{a}^{\dag} \hat{a} - \hat{b}^{\dag} \hat{b} \right)  \label{Sch}
\eea 
where $\hat{a}$ and $\hat{b}$ are the annihilation operators corresponding to the optical paths inside the interferometer, as labeled in Fig. \ref{a}. The operators defined in (\ref{Sch}) satisfy the SU$(2)$ algebra, hence $[\hat{J}_i, \hat{J}_j]=i \epsilon_{ijk} \hat{J}_{k}$. It will also be useful to define 
\beq
\hat{J}_0 = \frac{1}{2} \left( \hat{a}^{\dag} \hat{a} + \hat{b}^{\dag} \hat{b} \right),
\eeq
which commutes with all the operators in (\ref{Sch}).

For an arbitrary state $\ket{\Psi}$ created inside the interferometer, we can evaluate (\ref{QFI}). Using $\ket{\Psi_\varphi} = \hat{U} \ket{\Psi}$ and rewriting $\hat{U}$ as $\exp( i\hat{J}_0 \phi) \exp(i \hat{J}_3 \varphi)$, where $\phi = \varphi_a + \varphi_b$, we obtain that $\ket{\Psi_\varphi^{\prime}} = \hat{U}^{\prime}\ket{\Psi} = i \hat{J}_{3}\ket{\Psi_\varphi}$. Therefore, equations (\ref{QFI}) and (\ref{CRB}) give
\beq
\Delta^2 \varphi \ge \frac{1}{4 \Delta^2 J_3}, \label{CR}
\eeq
where $\Delta^2 J_3 = \bra{\Psi} \hat{J}_3^2 \ket{\Psi} - \bra{\Psi} \hat{J}_3 \ket{\Psi}^2$.

\section{Bell state amplification strategy}

\subsection{Probe preparation}
Now we are in a position to describe our protocol and demonstrate its usefulness for phase estimation. We start by injecting a single photon state $\ket{1}$  into the primary port of the initial 50:50 BS and the vacuum $\ket{0}$ into the secondary port, thus, generating the path entangled state 
\beq
\ket{1001}=\frac{1}{\sqrt{2} }\left(\ket{1,0}+i\ket{0,1} \right). \label{bell}
\eeq
Here we are using the notation $\ket{m,n} \equiv \ket{m}_a \otimes \ket{n}_b$, indicating that there are $m$ photons in mode $\hat{a}$ and $n$ photons in mode $\hat{b}$. This is a single photon noon state that possesses the properties of entanglement and coherence necessary for quantum-enhanced metrology. We amplify this state by performing the squeezing operation 
\beq
\hat{S}_{c}(r, \theta) = e^{\frac{r}{2} ( e^{-i \theta} \hat{c}^2 - e^{i \theta} \hat{c}^{\dag 2})}
\eeq
on each mode, where $\hat{c}$ is either $\hat{a}$ or $\hat{b}$ (Fig. \ref{b}). In practice this can be accomplished using type-I parametric down-conversion. The resulting probe state is
\bea
\ket{\Psi} &=& \frac{\hat{S}_{a} \otimes \hat{S}_{b}}{\sqrt{2} }\left(\ket{1,0}+ i\ket{0,1} \right) \nonumber \\ 
&=& \frac{1}{\sqrt{2}} \left( \ket{\Phi_1}\ket{\Phi_0} + i\ket{\Phi_0}\ket{\Phi_1} \right), \label{probe}
\eea    
where $\ket{\Phi_i} = \hat{S}_{c} \ket{i}$ and $i=0$ or $1$. Furthermore, 
\beq
\ket{\Phi_0} = \sum_{n=0}^{\infty} C_{n}\ket{2n},  \label{S0}
\eeq
and 
\beq
\ket{\Phi_1} = \text{sech} \, {r} \sum_{m=0}^{\infty} \sqrt{2m+1} C_{m}\ket{2m+1},  \label{S1}
\eeq
where $C_{n}= \sqrt{\text{sech} \, {r}} {\sqrt{(2n)!}}(-\tanh{r})^{n}/(2^n n!)$ \cite{PhysRevD.32.400}. Without loss of generality, we set $\theta=0$. 

\subsection{Demonstrating supersensitivity and super-resolution}

\begin{figure}[t]
\centering
\includegraphics[width=7.5 cm]{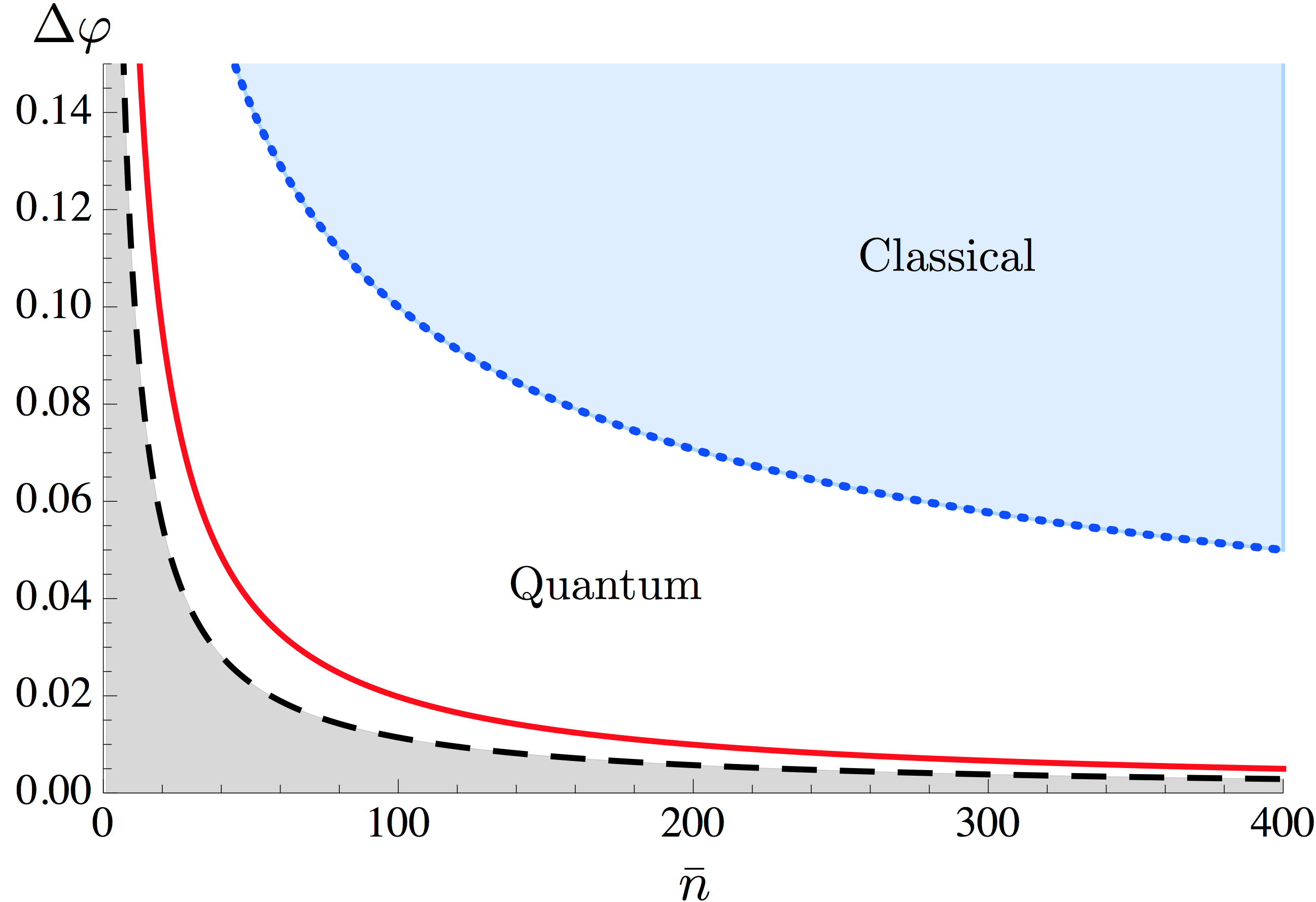}
\caption{(Color online)  Sensitivity of phase estimation versus mean photon number as given by the Cram\'{e}r-Rao bound (dashed, black line). The shaded gray region is inaccessible by any parameter estimation procedure which employs (\ref{probe}). Sensitivity of phase estimation, when $\varphi$ is an integer multiple of $\pi$, obtained from parity measurements (solid, red line) and the shot-noise limit (dotted, blue line) are also depicted. The white region is only accessible to quantum phase estimation protocols that employ non-classical correlations.}
\label{Min}
\end{figure}

For a given squeezing strength $r$, the prepared probe (\ref{probe}) has mean photon number $\bar{n} = \bra{\Psi}( \hat{a}^{\dag}\hat{a} + \hat{b}^{\dag}\hat{b} )\ket{\Psi}= 1+4 \sinh^2{r}$. Since this is a monotonic function, we may describe squeezing strength with either $r$ or $\bar{n}$. Finally, the Cram\'{e}r-Rao bound for this scheme is obtained using (\ref{CR}) and (\ref{probe}):
\beq
\Delta^2 \varphi \ge \frac{4}{\left(3 \bar{n}^2+6 \bar{n}-5\right)}.
\eeq
See Fig. \ref{Min} for a plot of this lower bound on $\Delta \varphi$. 

\begin{figure}[t]
\centering
\includegraphics[width=7 cm]{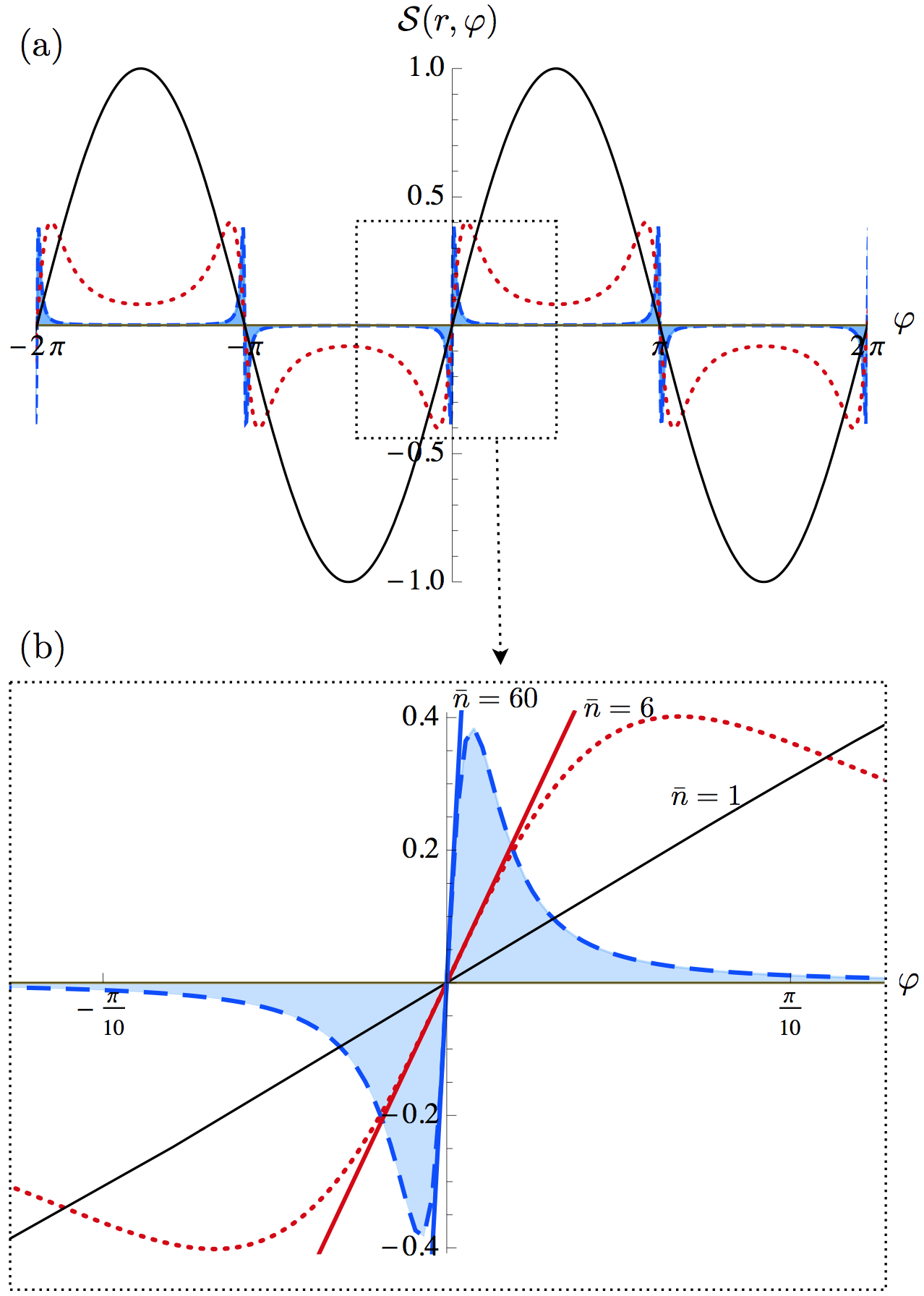}
\caption{(Color online) (a) Plot of $\mathcal{S}(r,\varphi)$ versus $\varphi$ over two periods. The signal $\mathcal{S}(r,\varphi)$ is shown for different squeezing strengths, corresponding to $\bar{n}=1$ (solid, black line), $\bar{n}=6$ (dotted, red line), and $\bar{n}=60$  (dashed, blue line). (b) Magnification of (a) near the origin. The linear approximations $(\bar{n}+1)\varphi/{2}$ are plotted as solid lines.}
\label{Sig}
\end{figure} 

This demonstrates the supersensitivity of the probe (\ref{probe}) to changes in $\varphi$. However, to illustrate the usefulness of (\ref{probe}) in a phase estimation protocol we must provide a measurement that can be performed in order to determine $\varphi$ with sub-shot-noise error. We accomplish this task by first passing the modified probe state $\ket{\Psi_{\varphi}}=\hat{U}\ket{\Psi}$ through a second 50:50 BS. Then, we perform a parity measurement on one of the output optical paths. This measurement was introduced into optical interferometry by Gerry and Campos \cite{PhysRevA.61.043811, PhysRevA.64.063814}, and it corresponds to determining the evenness or oddness of the photon number in a given beam. The effectiveness of parity detection in quantum metrology has been shown for a wide range of probe states \cite{PhysRevA.87.043833, doi:10.1080/09500340.2011.585251, PhysRevLett.104.103602}. The Hermitian operator corresponding such a measurement on mode $\hat{b}$ is 
\beq
\hat{\Pi} = \left( -1 \right)^ {\hat{b}^{\dag} \hat{b}} = e^{i \pi \left(\hat{J}_0 -\hat{J}_3 \right)} \label{parity}.
\eeq    
Note that the final state, right before making this measurement, is $\ket{\Psi_{f}} \equiv \hat{B} \hat{U} \ket{\Psi}$. Therefore, the expectation value of the measurement outcome is given as
\bea
&\, & \langle \hat{\Pi} \rangle_{\varphi} = \bra{\Psi} \hat{U}^{\dag} \hat{B}^{\dag} \hat{\Pi}   \hat{B} \hat{U} \ket{\Psi} \nonumber \\
&=& \bra{\Psi} e^{i \pi \hat{J}_0} e^{-i \varphi \hat{J}_3} e^{-i\frac{\pi}{2} \hat{J}_1} e^{-i \pi \hat{J}_3} e^{i \frac{\pi}{2} \hat{J}_1} e^{i \varphi \hat{J}_3} \ket{\Psi} \nonumber \\
&=&  \bra{\Psi} e^{i \pi \hat{J}_0} e^{-i \varphi \hat{J}_3} e^{i \pi \hat{J}_2} e^{i \varphi \hat{J}_3} \ket{\Psi}. \label{exp}
\eea   
In the previous step we have used the fact that $\hat{J}_0$ commutes with all $\hat{J}_i$ operators and the following relations from Ref. \cite{PhysRevA.66.013804} 
\bea
e^{-i\frac{\pi}{2} \hat{J}_1} \hat{J}_3 e^{i \frac{\pi}{2} \hat{J}_1} &=& -\hat{J}_2, \nonumber \\
e^{-i\frac{\pi}{2} \hat{J}_1} e^{-i \pi \hat{J}_3} e^{i \frac{\pi}{2} \hat{J}_1} &=& e^{i \pi \hat{J}_2}.
\eea
Using equations (\ref{probe}) to ( \ref{S1}) and (\ref{exp}), we find
\bea
\mathcal{S}(r,\varphi) &\equiv& \langle \hat{\Pi} \rangle_{\varphi + \frac{\pi}{2}} \nonumber \\
&=& \frac{ \sin{\varphi} \cosh (2 r) \, \text{sech}^6 \, r }{\left( 1- 2 \cos(2 \varphi )  \tanh^2{r} +  \tanh^4 {r} \right)^{3/2}},\label{signal}
\eea
where we have introduced an additional $\pi/2$ phase shift in order to simplify our analysis. Also, following \cite{PhysRevA.66.013804} we have used the relation $\exp(i \pi \hat{J}_2) \ket{n}_a \ket{m}_b$ $= (-1)^n \ket{m}_a \ket{n_b}$ from \cite{PhysRevA.40.1371} in deriving Eq. (\ref{signal}). 

The expectation value of the parity operator $\mathcal{S}(r,\varphi)$ represents the measurement signal from which we deduce the value of $\varphi$. As depicted in Fig. \ref{Sig}, this is a $2\pi$ periodic function that changes rapidly near $\varphi= k \pi$ ($k$ being any integer) when the squeezing strength $r$ is increased from 0. More specifically, when $r=0$ we get the signal for the single photon case, which is simply $\sin{\varphi}$; moreover, increasing $r$ leads to an increased magnitude of slope (i.e. increased $|{\partial \mathcal{S}(r,\varphi)}/{\partial \varphi}|$) around $\varphi=k \pi$. Therefore, we say that the measurement signal is super-resolving at these values. This can be seen from the fact that $\mathcal{S}(r,\varphi)$ is approximated by $({\bar{n}+1})(\varphi-k \pi)/2$ to second order in $(\varphi - k \pi)$ at $\varphi=k \pi$, i.e.,
\beq
\mathcal{S}(r,\varphi)= \left(\frac{\bar{n}+1}{2}\right) \left( \varphi - k \pi \right) + O\left( \left(\varphi - k \pi \right)^3 \right).
\eeq
This is depicted for $\varphi=0$ in Fig. \ref{Sig}, where we have plotted $\mathcal{S}(r,\varphi)$ for multiple values of $r$ (corresponding to $\bar{n}=1,6$ and $60$) along with the linear approximations $({\bar{n}+1})\varphi/{2}$. 

Since the measurement signal $\mathcal{S}(r,\varphi)$ is approximately linear near $\varphi= k \pi$, the mean-squared error of $\varphi$ at these values can be obtained using the linear propagation of error relation (\ref{error}):    
\beq
\Delta \varphi =  \frac{\sqrt{1 -  \mathcal{S}\left(r,k \pi \right)^2}}{|\left[{\partial \mathcal{S}\left(r,\varphi \right) }/{\partial \varphi} \right]_{\varphi=k \pi}|} = \frac{2}{\bar{n}+1}. \label{errorf}
\eeq 
This is a substantial improvement over the shot-noise limit (Fig. \ref{Min}). There are quantum metrology schemes that utilize the parity measurement and have better error scaling than (\ref{errorf}). This includes noon state metrology which beats the shot-noise limit by a factor of $\sqrt{\bar{n}}$. However, the difficulties in experimentally realizing these states inhibits their implementation into practical metrology protocols \cite{matthews2013practical}. Another quantum MZI scheme that gives better scaling than (\ref{errorf}) employs a two mode squeezed vacuum (TMSV) state \cite{PhysRevLett.104.103602}, however, the highest observed two mode squeezing to date (10 dB)  \cite{PhysRevLett.100.033602} would result in a TMSV state with $\bar{n} \approx 2$. Therefore, although these schemes provide greater sensitivity than (\ref{errorf}), they do not lead to a meaningful improvement over classical metrology protocols due to their low intensity (small $\bar{n}$). In contrast, our protocol provides a way to overcome the challenge of engineering high-intensity probe states for quantum-enhanced interferometry. In fact, a state similar to (\ref{probe}) was created by De Martini \emph{et al.} for the purpose of studying micro-macro entanglement \cite{PhysRevLett.100.253601}. In this experiment only one of the spatial modes of (\ref{bell}) was squeezed using an optical parametric amplifier and a value of $\bar{n} \approx 3.5 \times 10^4$ was reported for the resulting state. 

Although the resolution resulting from a probe state containing on average approximately $10^{-4}$ is still significantly lower than the resolution of current laser interferometry, the presented scheme may be useful for practical interferometry in the near future. The effect of photon loss on Bell state amplification by squeezing has been studied in the context of micro-macro entanglement verification \cite{PhysRevLett.110.170406, PhysRevLett.108.120404}. In order to demonstrate the usefulness of this scheme in the presence of noise, a similar analysis must be carried out for metrology \footnote{In preparation}.          

Like the noon state and TMSV state schemes, the phase information is extracted in our protocol by implementing a parity measurement, which can be done using a photon number resolving detector. Current technologies can resolve tens of photons at a time with an efficiency of $95\%$ \cite{miller:791, Lita:08, PhysRevA.80.043822, PhysRevLett.96.203601}. Knowledge of exact photon number represents more information than knowing the parity of the number of photons. Therefore, the way ahead may be to perform parity detection without determining the exact photon number, which is a currently active field of research \cite{PhysRevA.66.013804,PhysRevLett.104.103602, PhysRevA.72.053818}. 

\section{Conclusion}

In conclusion, we have outlined a phase estimation protocol for optical interferometry in which the probe state is prepared by amplifying a path entangled bell state to a macroscopically entangled state. When a parity measurement is utilized, this metrology strategy is super-resolving and supersensitive, beating the shot-noise limit by a factor greater than $\sqrt{\bar{n}}$/2. The macroscopically entangled probe state can be produced with current technological capabilities with $\bar{n}$ in the order of many thousands of photons.

This work was funded by NSERC. The authors would like to thank C. Simon and S. Raeisi for useful discussions.

\bibliography{PhD}

\end{document}